\documentclass[final,times,twocolumn]{elsarticle}

\usepackage{amssymb}

\usepackage{url}

\usepackage[usenames,dvipsnames]{xcolor}

\journal{SoftwareX}

\begin{document}

\begin{frontmatter}



\title{Qudi: a modular python suite for experiment control and data processing}


\author[ulm]{Jan M Binder}
\author[ulm,copenhagen]{Alexander Stark}
\author[ulm]{Nikolas Tomek}
\author[ulm]{Jochen Scheuer}
\author[ulm]{Florian Frank}
\author[ulm]{Kay D Jahnke}
\author[ulm]{Christoph M\"uller}
\author[ulm]{Simon Schmitt}
\author[ulm]{Mathias H Metsch}
\author[ulm]{Thomas Unden}
\author[copenhagen]{Tobias Gehring}
\author[copenhagen]{Alexander Huck}
\author[copenhagen]{Ulrik L Andersen}
\author[ulm]{Lachlan J Rogers\footnote{Corresponding author: lachlan.j.rogers@quantum.diamonds}}
\author[ulm,iqst]{Fedor Jelezko}

\address[ulm]{Institute for Quantum Optics, Ulm University, Albert-Einstein-Allee 11, Ulm 89081, Germany}
\address[copenhagen]{Department of Physics, Technical University of Denmark, Fysikvej, Kongens Lyngby 2800, Denmark}
\address[iqst]{Center for Integrated Quantum Science and Technology (IQ$^\mathrm{{st}}$), Ulm University, 89081 Germany}

\begin{abstract}
Qudi is a general, modular, multi-operating system suite written in Python 3 for controlling laboratory experiments.
It provides a structured environment by separating functionality into hardware abstraction, experiment logic and user interface layers.
The core feature set comprises a graphical user interface, live data visualization, distributed execution over networks, rapid prototyping via Jupyter notebooks, configuration management, and data recording.
Currently, the included modules are focused on confocal microscopy, quantum optics and quantum information experiments, but an expansion into other fields is possible and encouraged.
Qudi is available from \url{https://github.com/Ulm-IQO/qudi} and is freely useable under the GNU General Public Licence.
\end{abstract}

\end{frontmatter}


\section{Motivation and significance}

Modern scientific experiments typically rely on multiple hardware devices working together in a coordinated fashion.
In many instances, the hardware devices are commercial products with programming interfaces for direct control via custom software.
The unique combination of such devices is then specific to a given experiment.
Efficient control of such experiments requires software that is capable of coordinating the operation of multiple devices.
In addition, data interpretation is facilitated by rapid data processing and visualization.

These challenges are exemplified when studying color centers in diamond as solid state quantum emitters for sensing, spin manipulation and quantum information technologies.
It is typical for such experiments to be performed on a ``home-built confocal microscope'' \cite{jelezko2004observation, balasubramanian2009ultralong, fedder2011towards, waldherr2011dark, dolde2014high-fidelity}.
As evidenced by the 2014 Nobel Prize in Chemistry, these techniques have expanded beyond the context of physics and now this kind of microscope is pushing advances
in biology \cite{rocker2009quantitative, grotjohann2011diffraction-unlimited, gottfert2013coaligned} and nanotechnology \cite{harke2008resolution, hell2009diffraction-unlimited}.
A wide range of hardware is used for such experiments, but there is a paucity of mature and flexible lab control software to operate the aparatus.

Here, we present Qudi, a Python software suite for controlling complex experiments and managing the acquisition and processing of measurement data. 
Despite being developed in the context of quantum optics laboratories, the core Qudi framework is broadly applicable to many scenarios involving coordinated operation of multiple experiment devices.
The free and open-source nature of Qudi makes it possible for anyone to use and modify the software to fit their research needs,  and the modular code design simplifies this task.
Qudi continues to be actively developed, but it is already mature enough for reliable laboratory use \cite{jantzen2016nanodiamonds}.

\section{Software description}

\subsection{Why Python?}

Python was chosen as the programming language for Qudi because of its conceptual synergy with the goals of the project.
As a dynamic, strongly typed, scripting language, Python has become a popular choice for scientific programming \cite{perkel_programming_2015, cass_2015_programming_language} as the importance of scientific software increases \cite{singh_chawla_unsung_2016}.
Python's high level of abstraction makes it human-readable and concise, providing a direct advantage for laboratory programming typically performed by scientists rather than dedicated software developers.
Source code availability under an open-source license, the built-in modular structure of Python and good community support lower the initial hurdle to learn the language.
Additionally, most laboratory hardware has at least an application programming interface (API) specified for the C programming language, which can be accessed by Python.

Scripting languages cannot replace established compiled programming languages for tasks where processing performance or memory efficiency is required
but they are very useful to glue together different components in order to benefit from the advantages each of them can offer \cite{ousterhout_scripting_1998}.
This is closely aligned with the concept of Qudi ``gluing'' together various devices and control methods for specific complex experiments.

\subsection{Qudi design}
\begin{figure*}
	\centering
		\includegraphics[width=\textwidth]{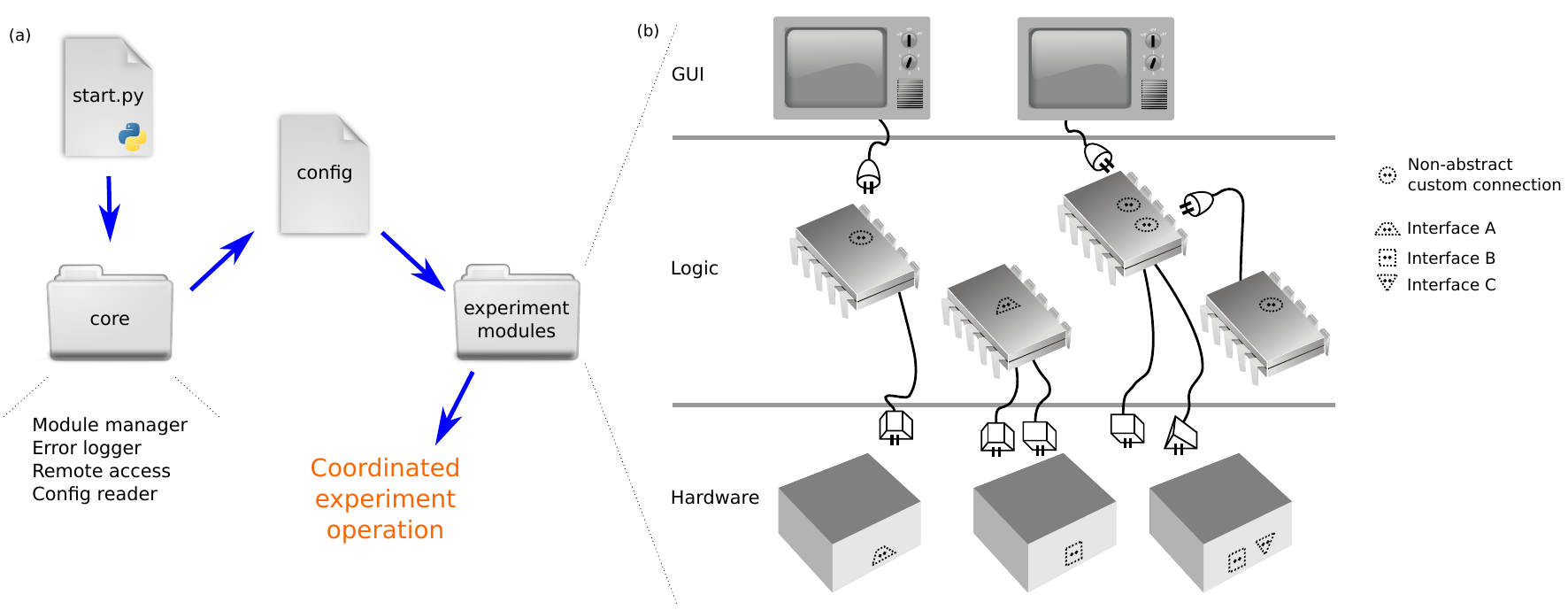}
	\caption{
		Qudi functional and structural design.
		(a) A user launches Qudi by running the start.py file, which loads the core components
        that take care of configuration parsing, module management, error logging, and remote network access.
		The module manager reads a configuration file to determine how a set of experiment modules should be configured for the specific laboratory aparatus.
        Hence, the experiement can be carried out.
		(b)
		There is a strongly-enforced three-layer design for all Qudi experiment modules.
		Specific measurements are written as logic modules, including the required tasks and data analysis.
		These logic modules connect down to hardware modules via well-defined interfaces, meaning that the experiment 
        itself is hardware agnostic as long as the hardware can fulfill the minimum requirements.
		GUI modules connect to the logic and provide a way for a user to operate the experiment, as well as a means to display data and calculated results.
		The careful separation of the GUI from the logic means that it is equally easy to operate experiments in a ``headless'' scripted manner.
	}
	\label{fig:Qudi_structure_opt1}
\end{figure*}

The Qudi suite consists of a collection of modules that are loaded and connected together by a manager component according to settings given in a configuration file as shown in Fig. \ref{fig:Qudi_structure_opt1}(a).
Startup is initiated by a single executable python file, and the manager component provides core functions for logging, error handling, configuration reading, and remote access.
Additionally, the manager also administers the other modules by providing functionality for module loading, module dependency resolution and connection, concurrent execution and network access to modules running on other computers.
This core infrastructure makes it easier to rapidly develop modules for new experiments by providing structure and starting points.
The program startup code, manager and logging components were initially derived from similar elements contained within the neurophysiology software ACQ4 \cite{campagnola_2014_acq4}.

A typical Qudi session will proceed as follows:
On startup, the supervisor process, for example an IDE, creates a Qudi process.
In this Qudi process, the manager component reads the configuration file, sets up the log file and loads the modules designated in the startup section of the configuration file.
Typically, the startup section will -- but does not have to -- contain at least the Manager GUI and the tray icon module.
The user can then load the modules specified in the configuration file for the desired measurement from this GUI or a Jupyter notebook to perform the measurement.
Some of the science modules in Qudi were inspired by the pi3diamond software~\cite{
fedder2011towards, 
waldherr2011dark,
dolde2014high-fidelity,
pi3diamond2016source, 
dolde2011electric-field, 
michl2014perfect}.

The modules that relate to particular experiments are divided into three categories: hardware interaction, experiment ``logic'', and user interface.
These categories and the relationships between them are illustrated in Fig. \ref{fig:Qudi_structure_opt1}(b).
The division into hardware, logic, and interface represents a clear separation of tasks that improves reliability and flexibility of the Qudi code.
It also simplifies the implementation of new experiment modules.
The fundamental three-fold distinction is at the basis of Qudi's adaptability, and makes Qudi an experiment control software in contrast to a general software framework.

\subsubsection {Logic modules}

Logic modules control and synchronize a given experiment.
They pass input parameters from the user interface to the respective hardware modules, and process measurement data in the desired way.
These modules control the information exchange between different hardware modules and perform all necessary computations and conversions.

Logic modules are the only type of modules that are allowed to interact with each other.
They are also the only type of module that has its own thread and event loop.
Therefore they are the place where concurrent execution of tasks and synchronization of different devices is handled.
All steps from the start of a measurement to its end, including data evaluation and storage are performed by the logic.
This goes as far as producing ``publication ready'' plots of data that are saved together with the raw data and which provide a good overview or can be sent to collaborators without post-processing.

\subsubsection{Hardware abstraction via interfaces}

Today it is possible and even necessary to control most experiment hardware remotely.
Unfortunately, the command structure, grammar, measurement units and connection methods differ widely between device models or devices from different suppliers of experiment hardware.
To get the most re-usability out of logic modules, it must be possible to interchange hardware modules for measurement devices that provide 
similar functionality, but work and communicate differently.
It is the task of the hardware modules to overcome these problems by translating the commands given by the logic into the ``language'' of the specific hardware.

The problem is solved by defining an interface, a set of functions that a hardware module of a given type must implement, in order to make a certain measurement work.
This set of functions is defined in a class (named ...Interface in a file in the interface folder) where the default implementation of each function raises an exception, if it is not replaced in the device-specific implementation.
This class is then inherited by the actual implementing hardware module and all inherited functions must be overwritten.

Hardware modules can represent virtual dummy or mock hardware, which emulates the functionality of a device. 
Those dummies could load recorded measurement files, create arbitrary data or may perform real physical simulations of measurements, where the result is prepared according to the interface commands which the logic can access.
One of the most significant uses of dummy hardware modules is to test the experiment logic without being connected to any actual hardware.


\subsubsection{Advanced abstraction via ``interfuses''}

Building on the abstraction of interfaces, Qudi introduces an additional concept to facilitate the reuse of modules.
This ability is provided by interfuse modules which  interconnect (or fuse) different hardware or logic modules to modify their interface behavior or to achieve a task for which these modules were not originally designed.
%

%
An interfuse is a logic module that implements a hardware interface.
In doing so, it pretends to be hardware that can connect to an experiment logic module.
This allows the core experiment functions to remain in the logic module, while altering the kind of data that is measured.
%
%
%
A tangible example helps clarify this concept.
A confocal image (2D array) can represent single fluorescence values from a photon counter for each position (x, y).
An interfuse makes it possible to replace the counter data with spectrometer measurements at each pixel, allowing fluorescence to be imaged with arbitrary spectral filtering.
This practice improves maintainability and prevents code duplication.

%
%
%
%

The other reason to use interfuses is where a desired feature would require altering an existing interface definition.
%
%
For example, an interfuse can perform the coordinate transform to correct for a tilted sample  in a confocal scan.
As a result, the tilted surface appears flat in the confocal image and can then be imaged at a consistent depth.  
%


\subsubsection{GUI}

Qudi GUI modules create windows on the screen that a user can interact with, allowing experiment control and data visualisation.
Their purpose is to offer a convenient way for the user to interact with logic modules, however Qudi is fully functional without the GUI modules.
The logic can also be controlled by the integrated IPython console or from a Jupyter notebook.
For this reason, GUI modules are not allowed to interact with each other or the hardware directly and they do no data processing.

The Qudi graphical user interface (GUI) is built with Qt \cite{qt-software}, offering users a familiar appearance.
Qt is suitable due to its multi-platform GUI toolkit that provides good Python bindings \cite{pyqt-software,pyside-software} and makes it possible to separate the GUI design from the implemented functionality.
Also, Qt's multi-thread ability ensures good scalability and parallel processing, which are essential requirements for complex experiments.
Furthermore, Qt implements a signal-slot mechanism \cite{qt-software_signal_slot} that is very useful for concurrency, modular design, and interaction between GUI modules and logic modules.
On top of this, the Python library PyQtGraph \cite{pyqtgraph-software} makes it easy to create interactive, frequently updated 2- and 3-dimensional plots.
%

The user interface can be edited graphically in Qt Designer and is stored as an XML file.  For rapid prototyping, this file can be (re)loaded by a running Python program.
The GUI design strives to adhere to the KDE Human Interface Guidelines \cite{KDEHIG}, as these stress the importance of interface familiarity
and they work well with the default set of Qt user interface elements.

\subsubsection{Interactive Scripting}

Interactive scripting provides a powerful additional user-interface for a flexible software suite. 
Qudi contains a built-in console with a fully integrated IPython interpreter.
In addition, Qudi can be controlled from a Jupyter Notebook.
This makes it possible to write a scripted document with incremental execution as well as inline visualisation and analysis.
%
%
Both the console and the Jupyter notebook can control all of the internal states of the Qudi software.
These features enable rapid experiment prototyping, since a developer can test different approaches before commiting to changes in hardware or logic modules.

\section{Impact and reuse potential}

The Qudi suite is useful for any small to medium-size computer-controlled laboratory experiment.
Its modular design combined with the use of interface definitions makes it easy to integrate new hardware into an existing experiment.
Moreover, this design offers the capability to easily reuse existing modules in new experiments.
The Qudi core infrastructure is broadly applicable, even beyond the context of confocal microscopy or physics experiments in general.

Qudi is of more tangible impact to the quantum optics community in particular.
The existing modules already offer control over
confocal microscopes, 
electromagnets,
motorised stages,
lasers,
(arbitrary) signal generators,
and other devices used in this field of research.
Furthermore, typical measurement protocols and data analysis functions are already implemented.
These existing modules make Qudi a ready-to-use Python-based software suite for quantum optics labs, independent from the individual hardware and measurement schemes used by different groups.

\section{Illustrative example}

One example which highlights the convenience of Qudi is the measurement of optically detected magnetic resonance (ODMR) on single color centers in diamond \cite{gruber_scanning_1997,jelezko_read-out_2004}.
Such an experiment requires the coordinated operation of a scanning confocal microscope and a microwave source.
This section describes how such a measurement is performed with Qudi.

\begin{figure*}
  \centering
  \includegraphics[width=\textwidth]{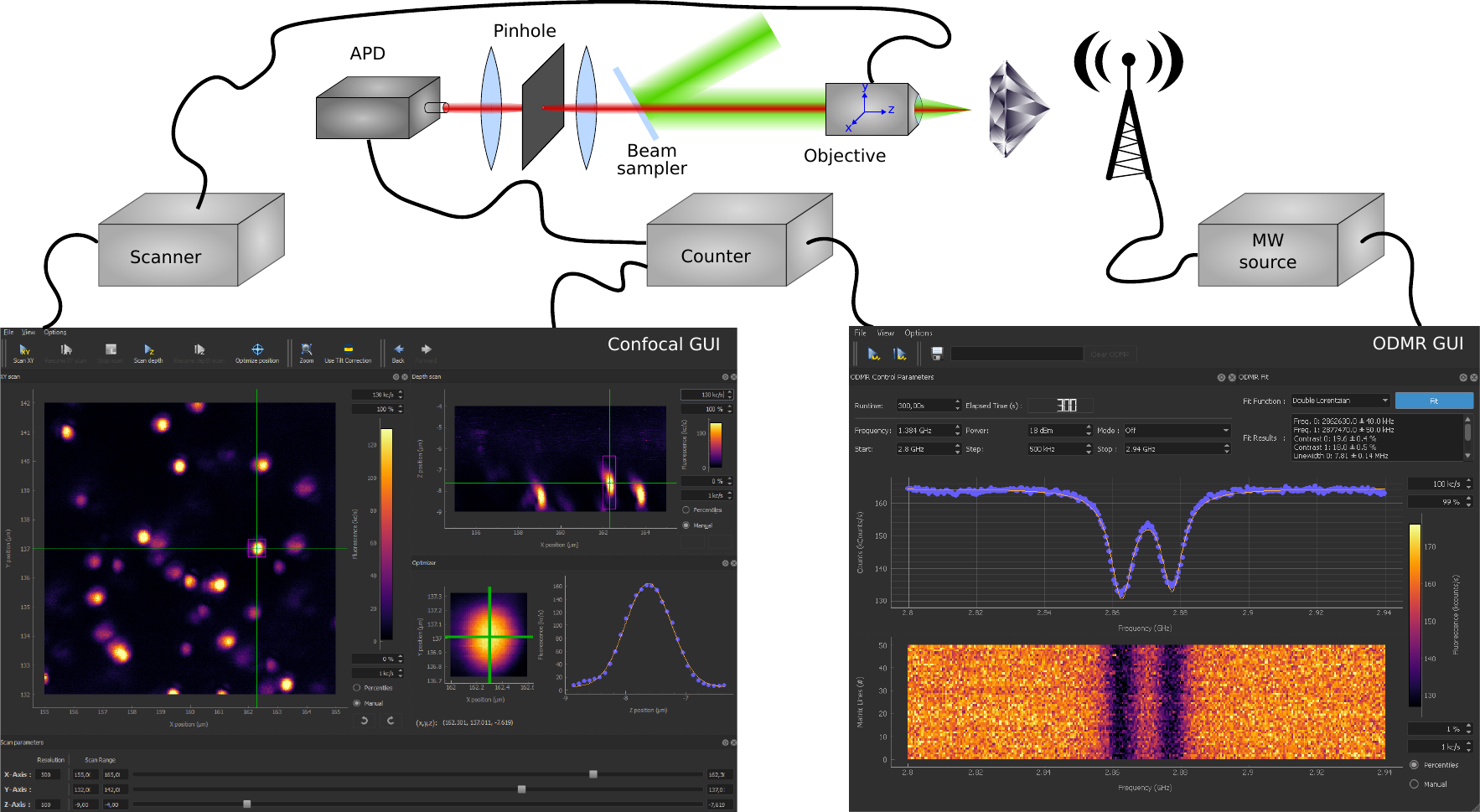}
  \caption{%
	  Simplified illustration of Qudi used to perform ODMR experiments in the laboratory. 
	  The experimental setup consists of three main parts. 
	  The confocal microscope is used to image the red fluorescence of color centers in diamond using a green excitation laser. 
	  The objective can be scanned in all three dimensions by scanner hardware. 
	  An avalanch photodiode (APD) detects red fluorescence photons which are counted by digital data acquisition hardware. 
	  In addition, a signal generator exposes the color centers to a microwave field which lowers the fluorescence at certain resonance frequencies (ODMR). 
	  The Confocal GUI shows the fluorescence images used to position the optical focal spot and the ODMR GUI displays the microwave resonance spectra.
	  %
	  %
	  This figure illustrates the experience of a user, and does not show the logic modules which perform experimental functions.
	  %
  }%
    \label{fig:illustrative_example}
\end{figure*}

The first step is to find a single color center inside the diamond. 
The Qudi confocal GUI and logic modules are used to move a diffraction-limited focal spot through the diamond sample in three dimensions \cite{pawley1996handbook,beveratos2001nonclassical,sipahigil2014indistinguishable,hausler_optical_2014}.
This is achieved by scanning hardware that is controlled by the confocal logic.
A photon counter records the fluorescence measured by the confocal microscope, and this hardware device sends the data to the confocal logic.
The confocal logic produces an image of fluorescence as a function of position, and the GUI presents this image to the user.
Figure \ref{fig:illustrative_example} shows the confocal GUI with an x-y image on the left and an x-z image on the right.

%
%
%

%
%
In order to focus on a single center, the user places the confocal cursor near a promising spot.
An optimizer module performs a series of close-range scans around the cursor, 
and the optimal position of maximum fluorescence is found via a fitting module built on the lmfit package \cite{matt_newville_2016_58759}. 
%
%
A 2D gaussian fit is performed on the x-y plane scan  and for the third dimension a 1D gaussian fit on the z line scan.
These are shown in Figure \ref{fig:illustrative_example} on the lower right of the Confocal GUI.
Finally, the optimizer module moves the scanning hardware to the optimal position focussed on the desired single color center.

In addition to spatial alignment, a microwave resonance condition has to be matched in order to detect the desired change in optical signal \cite{gruber_scanning_1997,jelezko_read-out_2004,nizovtsev2010quantum,london_detecting_2013,dolde2013room}.
%
%
The ODMR logic module controls the frequency of a microwave source while recording the fluorescence level.
The design of Qudi means that the ODMR logic is easily capable of driving a variety of microwave source hardware, increasing flexibility in the laboratory.
%
%

An ODMR experiment is performed by sweeping the microwave frequency and recording the fluorescence.
Recorded data are shown live on screen in the ODMR GUI as both the fluorescence sum of all frequency sweeps and as a matrix plot containing each sweep (Fig. \ref{fig:illustrative_example}, lower right).
ODMR scans of several spots can be measured automatically by saving the color center positions and then using a script to move from spot to spot, optimizing the position on each site and recording an ODMR spectrum.

\section{Conclusions and future directions}

Qudi is a generally applicable experiment control software suite, with infrastructure to support modular design of experiments, signifcantly reducing the effort involved in constructing new experiments.
Qudi already offers a developed quantum-optics tool set capable of reliable laboratory operation, and a modern user interface.

There is continuing effort to expand the library of available science modules.
One priority for the future is to simplify the setup of Qudi by providing a graphical configuration editor.
Furthermore, it would be convenient to make Qudi installable from the Python Package Index.
In the context of experiment operation, enhanced automation capabilities are desired to allow a user to rearrange the existing functionality without programming.

\section*{Acknowledgements}
We would like to thank Boris Naydenov for advocating the use of a software platform that can be used
by the whole Quantum Optics institute and for maintaining the predecessor to this software.
Furthermore, we would like to thank Ou Wang, Gerhard Wolff, Samuel M\"uller and Andrea Kurz for contributions to the software, testing and reporting bugs.

This work was supported by the 
ERC, 
EU projects (SIQS, DIADEMS, EQUAM), 
DFG (FOR 1482, FOR 1493 and SFBTR 21), 
BMBF, 
the Volks\-wagen foundation,
the Innovation Foundation Denmark (EXMAD project no. 1311-00006B and Qubiz) and the Danish 
Research Council for Independent Research (DIMS project no. 4181-00505B, Individual Postdoc and Sapere Aude, 4184-00338B).


\appendix


\bibliographystyle{elsarticle-num}
\bibliography{paper_qudi}

\begin{thebibliography}{10}
\expandafter\ifx\csname url\endcsname\relax
  \def\url#1{\texttt{#1}}\fi
\expandafter\ifx\csname urlprefix\endcsname\relax\def\urlprefix{URL }\fi
\expandafter\ifx\csname href\endcsname\relax
  \def\href#1#2{#2} \def\path#1{#1}\fi

\bibitem{jelezko2004observation}
F.~Jelezko, T.~Gaebel, I.~Popa, A.~Gruber, J.~Wrachtrup, {Observation of
  {Coherent} {Oscillations} in a {Single} {Electron} {Spin}}, Phys. Rev. Lett.
  92~(7) (2004) 076401.
\newblock \href {http://dx.doi.org/10.1103/PhysRevLett.92.076401}
  {\path{doi:10.1103/PhysRevLett.92.076401}}.

\bibitem{balasubramanian2009ultralong}
G.~Balasubramanian, P.~Neumann, D.~Twitchen, M.~Markham, R.~Kolesov,
  N.~Mizuochi, J.~Isoya, J.~Achard, J.~Beck, J.~Tissler, V.~Jacques, P.~R.
  Hemmer, F.~Jelezko, J.~Wrachtrup, {Ultralong spin coherence time in
  isotopically engineered diamond}, Nature Materials 8 (2009) 383--387.
\newblock \href {http://dx.doi.org/10.1038/nmat2420}
  {\path{doi:10.1038/nmat2420}}.

\bibitem{fedder2011towards}
H.~Fedder, F.~Dolde, F.~Rempp, T.~Wolf, P.~Hemmer, F.~Jelezko, J.~Wrachtrup,
  \href{http://link.springer.com/article/10.1007/s00340-011-4408-4}{Towards
  {T}1-limited magnetic resonance imaging using {Rabi} beats}, Applied Physics
  B 102~(3) (2011) 497--502.
\newblock \href {http://dx.doi.org/10.1007/s00340-011-4408-4}
  {\path{doi:10.1007/s00340-011-4408-4}}.
\newline\urlprefix\url{http://link.springer.com/article/10.1007/s00340-011-4408-4}

\bibitem{waldherr2011dark}
G.~Waldherr, J.~Beck, M.~Steiner, P.~Neumann, A.~Gali, T.~Frauenheim,
  F.~Jelezko, J.~Wrachtrup,
  \href{http://link.aps.org/doi/10.1103/PhysRevLett.106.157601}{Dark {States}
  of {Single} {Nitrogen}-{Vacancy} {Centers} in {Diamond} {Unraveled} by
  {Single} {Shot} {NMR}}, Physical Review Letters 106~(15) (2011) 157601.
\newblock \href {http://dx.doi.org/10.1103/PhysRevLett.106.157601}
  {\path{doi:10.1103/PhysRevLett.106.157601}}.
\newline\urlprefix\url{http://link.aps.org/doi/10.1103/PhysRevLett.106.157601}

\bibitem{dolde2014high-fidelity}
F.~Dolde, V.~Bergholm, Y.~Wang, I.~Jakobi, B.~Naydenov, S.~Pezzagna, J.~Meijer,
  F.~Jelezko, P.~Neumann, T.~Schulte-Herbr{\"u}ggen, J.~Biamonte, J.~Wrachtrup,
  \href{http://www.nature.com/ncomms/2014/140228/ncomms4371/full/ncomms4371.html}{High-fidelity
  spin entanglement using optimal control}, Nature Communications 5 (2014)
  3371.
\newblock \href {http://dx.doi.org/10.1038/ncomms4371}
  {\path{doi:10.1038/ncomms4371}}.
\newline\urlprefix\url{http://www.nature.com/ncomms/2014/140228/ncomms4371/full/ncomms4371.html}

\bibitem{rocker2009quantitative}
C.~R{\"o}cker, M.~P{\"o}tzl, F.~Zhang, W.~J. Parak, G.~U. Nienhaus,
  \href{http://www.nature.com/nnano/journal/v4/n9/full/nnano.2009.195.html}{{A
  quantitative fluorescence study of protein monolayer formation on colloidal
  nanoparticles}}, Nature Nanotechnology 4~(9) (2009) 577--580.
\newblock \href {http://dx.doi.org/10.1038/nnano.2009.195}
  {\path{doi:10.1038/nnano.2009.195}}.
\newline\urlprefix\url{http://www.nature.com/nnano/journal/v4/n9/full/nnano.2009.195.html}

\bibitem{grotjohann2011diffraction-unlimited}
T.~Grotjohann, I.~Testa, M.~Leutenegger, H.~Bock, N.~T. Urban,
  F.~Lavoie-Cardinal, K.~I. Willig, C.~Eggeling, S.~Jakobs, S.~W. Hell,
  \href{http://www.nature.com/nature/journal/v478/n7368/full/nature10497.html}{{Diffraction-unlimited
  all-optical imaging and writing with a photochromic {GFP}}}, Nature
  478~(7368) (2011) 204--208.
\newblock \href {http://dx.doi.org/10.1038/nature10497}
  {\path{doi:10.1038/nature10497}}.
\newline\urlprefix\url{http://www.nature.com/nature/journal/v478/n7368/full/nature10497.html}

\bibitem{gottfert2013coaligned}
F.~G{\"o}ttfert, C.~A. Wurm, V.~Mueller, S.~Berning, V.~C. Cordes,
  A.~Honigmann, S.~W. Hell,
  \href{http://www.sciencedirect.com/science/article/pii/S0006349513006127}{Coaligned
  {Dual}-{Channel} {STED} {Nanoscopy} and {Molecular} {Diffusion} {Analysis} at
  20 nm {Resolution}}, Biophysical Journal 105~(1) (2013) L01--L03.
\newblock \href {http://dx.doi.org/10.1016/j.bpj.2013.05.029}
  {\path{doi:10.1016/j.bpj.2013.05.029}}.
\newline\urlprefix\url{http://www.sciencedirect.com/science/article/pii/S0006349513006127}

\bibitem{harke2008resolution}
B.~Harke, J.~Keller, C.~K. Ullal, V.~Westphal, A.~Sch{\"o}nle, S.~W. Hell,
  \href{http://www.osapublishing.org/abstract.cfm?uri=oe-16-6-4154}{Resolution
  scaling in {STED} microscopy}, Optics Express 16~(6) (2008) 4154--4162.
\newblock \href {http://dx.doi.org/10.1364/OE.16.004154}
  {\path{doi:10.1364/OE.16.004154}}.
\newline\urlprefix\url{http://www.osapublishing.org/abstract.cfm?uri=oe-16-6-4154}

\bibitem{hell2009diffraction-unlimited}
S.~W. Hell, R.~Schmidt, A.~Egner,
  \href{http://www.nature.com/nphoton/journal/v3/n7/full/nphoton.2009.112.html}{Diffraction-unlimited
  three-dimensional optical nanoscopy with opposing lenses}, Nature Photonics
  3~(7) (2009) 381--387.
\newblock \href {http://dx.doi.org/10.1038/nphoton.2009.112}
  {\path{doi:10.1038/nphoton.2009.112}}.
\newline\urlprefix\url{http://www.nature.com/nphoton/journal/v3/n7/full/nphoton.2009.112.html}

\bibitem{jantzen2016nanodiamonds}
U.~Jantzen, A.~B. Kurz, D.~S. Rudnicki, C.~Sch{\"a}fermeier, K.~D. Jahnke,
  U.~L. Andersen, V.~A. Davydov, V.~N. Agafonov, A.~Kubanek, L.~J. Rogers,
  F.~Jelezko, \href{http://arxiv.org/abs/1602.03391}{{Nanodiamonds carrying
  quantum emitters with almost lifetime-limited linewidths}}, arXiv:1602.03391
  [cond-mat, physics:physics, physics:quant-ph]ArXiv: 1602.03391.
\newline\urlprefix\url{http://arxiv.org/abs/1602.03391}

\bibitem{perkel_programming_2015}
J.~M. Perkel,
  \href{http://www.nature.com/doifinder/10.1038/518125a}{{Programming: {Pick}
  up {Python}}}, Nature 518~(7537) (2015) 125--126.
\newblock \href {http://dx.doi.org/10.1038/518125a}
  {\path{doi:10.1038/518125a}}.
\newline\urlprefix\url{http://www.nature.com/doifinder/10.1038/518125a}

\bibitem{cass_2015_programming_language}
S.~Cass,
  \href{http://spectrum.ieee.org/computing/software/the-2015-top-ten-programming-languages}{{The
  2015 {Top} {Ten} {Programming} {Languages}}} (Jul. 2015).
\newline\urlprefix\url{http://spectrum.ieee.org/computing/software/the-2015-top-ten-programming-languages}

\bibitem{singh_chawla_unsung_2016}
D.~{Singh Chawla}, \href{http://www.nature.com/doifinder/10.1038/529115a}{{The
  unsung heroes of scientific software}}, Nature 529~(7584) (2016) 115--116.
\newblock \href {http://dx.doi.org/10.1038/529115a}
  {\path{doi:10.1038/529115a}}.
\newline\urlprefix\url{http://www.nature.com/doifinder/10.1038/529115a}

\bibitem{ousterhout_scripting_1998}
J.~K. Ousterhout, {Scripting: higher level programming for the 21st {Century}},
  Computer 31~(3) (1998) 23--30.
\newblock \href {http://dx.doi.org/10.1109/2.660187}
  {\path{doi:10.1109/2.660187}}.

\bibitem{campagnola_2014_acq4}
L.~Campagnola, M.~B. Kratz, P.~B. Manis,
  \href{http://www.frontiersin.org/neuroinformatics/10.3389/fninf.2014.00003/abstract}{{ACQ4:
  an open-source software platform for data acquisition and analysis in
  neurophysiology research}}, Frontiers in Neuroinformatics 8~(3).
\newblock \href {http://dx.doi.org/10.3389/fninf.2014.00003}
  {\path{doi:10.3389/fninf.2014.00003}}.
\newline\urlprefix\url{http://www.frontiersin.org/neuroinformatics/10.3389/fninf.2014.00003/abstract}

\bibitem{pi3diamond2016source}
\href{https://github.com/HelmutFedder/pi3diamond}{pi3diamond source code}.
\newline\urlprefix\url{https://github.com/HelmutFedder/pi3diamond}

\bibitem{dolde2011electric-field}
F.~Dolde, H.~Fedder, M.~W. Doherty, T.~N{\"o}bauer, F.~Rempp,
  G.~Balasubramanian, T.~Wolf, F.~Reinhard, L.~C.~L. Hollenberg, F.~Jelezko,
  J.~Wrachtrup,
  \href{http://www.nature.com/nphys/journal/v7/n6/abs/nphys1969.html}{Electric-field
  sensing using single diamond spins}, Nature Physics 7~(6) (2011) 459--463.
\newblock \href {http://dx.doi.org/10.1038/nphys1969}
  {\path{doi:10.1038/nphys1969}}.
\newline\urlprefix\url{http://www.nature.com/nphys/journal/v7/n6/abs/nphys1969.html}

\bibitem{michl2014perfect}
J.~Michl, T.~Teraji, S.~Zaiser, I.~Jakobi, G.~Waldherr, F.~Dolde, P.~Neumann,
  M.~W. Doherty, N.~B. Manson, J.~Isoya, J.~Wrachtrup,
  \href{http://scitation.aip.org/content/aip/journal/apl/104/10/10.1063/1.4868128}{Perfect
  alignment and preferential orientation of nitrogen-vacancy centers during
  chemical vapor deposition diamond growth on (111) surfaces}, Applied Physics
  Letters 104~(10) (2014) 102407.
\newblock \href {http://dx.doi.org/10.1063/1.4868128}
  {\path{doi:10.1063/1.4868128}}.
\newline\urlprefix\url{http://scitation.aip.org/content/aip/journal/apl/104/10/10.1063/1.4868128}

\bibitem{qt-software}
{Qt | Cross-platform application development for desktop \& embedded},
  \url{https://www.qt.io/}, acessed 2016-08-07.

\bibitem{pyqt-software}
{Riverbank | Software | PyQt | What is PyQt?},
  \url{https://riverbankcomputing.com/software/pyqt/intro}, accessed
  2016-08-07.

\bibitem{pyside-software}
{PySide}, \url{https://wiki.qt.io/PySide}.

\bibitem{qt-software_signal_slot}
\href{http://doc.qt.io/qt-5/signalsandslots.html}{Signals \& {Slots} in {Qt}
  5}.
\newline\urlprefix\url{http://doc.qt.io/qt-5/signalsandslots.html}

\bibitem{pyqtgraph-software}
{PyQtGraph - Scientific Graphics and GUI Library for Python},
  \url{http://www.pyqtgraph.org/}, accessed 2016-08-07.

\bibitem{KDEHIG}
{KDE Human Interface Guidelines},
  \url{https://community.kde.org/index.php?title=KDE\_Visual\_Design\_Group/HIG\&oldid=72475},
  accessed 2016-10-11.

\bibitem{gruber_scanning_1997}
A.~Gruber, A.~Dr{\"a}benstedt, C.~Tietz, L.~Fleury, J.~Wrachtrup, C.~v.
  Borczyskowski,
  \href{http://www.sciencemag.org/content/276/5321/2012}{{Scanning {Confocal}
  {Optical} {Microscopy} and {Magnetic} {Resonance} on {Single} {Defect}
  {Centers}}}, Science 276~(5321) (1997) 2012--2014, 00662.
\newblock \href {http://dx.doi.org/10.1126/science.276.5321.2012}
  {\path{doi:10.1126/science.276.5321.2012}}.
\newline\urlprefix\url{http://www.sciencemag.org/content/276/5321/2012}

\bibitem{jelezko_read-out_2004}
F.~Jelezko, J.~Wrachtrup,
  \href{http://iopscience.iop.org/0953-8984/16/30/R03}{{Read-out of single
  spins by optical spectroscopy}}, Journal of Physics: Condensed Matter 16~(30)
  (2004) R1089, 00084.
\newblock \href {http://dx.doi.org/10.1088/0953-8984/16/30/R03}
  {\path{doi:10.1088/0953-8984/16/30/R03}}.
\newline\urlprefix\url{http://iopscience.iop.org/0953-8984/16/30/R03}

\bibitem{pawley1996handbook}
J.~Pawley, B.~R. Masters, {Handbook of biological confocal microscopy}, Optical
  Engineering 35~(9) (1996) 2765--2766.

\bibitem{beveratos2001nonclassical}
A.~Beveratos, R.~Brouri, T.~Gacoin, J.-P. Poizat, P.~Grangier, {Nonclassical
  radiation from diamond nanocrystals}, Physical Review A 64~(6) (2001) 061802.

\bibitem{sipahigil2014indistinguishable}
A.~Sipahigil, K.~D. Jahnke, L.~J. Rogers, T.~Teraji, J.~Isoya, A.~S. Zibrov,
  F.~Jelezko, M.~D. Lukin, {Indistinguishable photons from separated
  silicon-vacancy centers in diamond}, Physical review letters 113~(11) (2014)
  113602.

\bibitem{hausler_optical_2014}
A.~J. H{\"a}ussler, P.~Heller, L.~P. McGuinness, B.~Naydenov, F.~Jelezko,
  \href{https://www.osapublishing.org/oe/abstract.cfm?uri=oe-22-24-29986}{{Optical
  depth localization of nitrogen-vacancy centers in diamond with nanometer
  accuracy}}, Optics Express 22~(24) (2014) 29986.
\newblock \href {http://dx.doi.org/10.1364/OE.22.029986}
  {\path{doi:10.1364/OE.22.029986}}.
\newline\urlprefix\url{https://www.osapublishing.org/oe/abstract.cfm?uri=oe-22-24-29986}

\bibitem{matt_newville_2016_58759}
M.~Newville, A.~Nelson, A.~Ingargiola, T.~Stensitzki, D.~Allan, Michał, Glenn,
  Y.~Ram, MerlinSmiles, L.~Li, G.~Pasquevich, C.~Deil, D.~M. Fobes, Stuermer,
  T.~Spillane, ampolloreno, stonebig, P.~A. Brodtkorb, R.~Clarken,
  K.~Anagnostopoulos, A.~Almarza, B.~Gamari,
  \href{https://doi.org/10.5281/zenodo.58759}{lmfit-py 0.9.5} (Jul. 2016).
\newblock \href {http://dx.doi.org/10.5281/zenodo.58759}
  {\path{doi:10.5281/zenodo.58759}}.
\newline\urlprefix\url{https://doi.org/10.5281/zenodo.58759}

\bibitem{nizovtsev2010quantum}
A.~Nizovtsev, S.~Y. Kilin, P.~Neumann, F.~Jelezko, J.~Wrachtrup, {Quantum
  registers based on single NV+ n 13C centers in diamond: II. Spin
  characteristics of registers and spectra of optically detected magnetic
  resonance}, Optics and Spectroscopy 108~(2) (2010) 239--246.

\bibitem{london_detecting_2013}
P.~London, J.~Scheuer, J.-M. Cai, I.~Schwarz, A.~Retzker, M.~B. Plenio,
  M.~Katagiri, T.~Teraji, S.~Koizumi, J.~Isoya, R.~Fischer, L.~P. McGuinness,
  B.~Naydenov, F.~Jelezko,
  \href{http://link.aps.org/doi/10.1103/PhysRevLett.111.067601}{{Detecting and
  {Polarizing} {Nuclear} {Spins} with {Double} {Resonance} on a {Single}
  {Electron} {Spin}}}, Physical Review Letters 111~(6) (2013) 067601, 00011.
\newblock \href {http://dx.doi.org/10.1103/PhysRevLett.111.067601}
  {\path{doi:10.1103/PhysRevLett.111.067601}}.
\newline\urlprefix\url{http://link.aps.org/doi/10.1103/PhysRevLett.111.067601}

\bibitem{dolde2013room}
F.~Dolde, I.~Jakobi, B.~Naydenov, N.~Zhao, S.~Pezzagna, C.~Trautmann,
  J.~Meijer, P.~Neumann, F.~Jelezko, J.~Wrachtrup, {Room-temperature
  entanglement between single defect spins in diamond}, Nature Physics 9~(3)
  (2013) 139--143.

\end{thebibliography}

\end{document}